# Routing Protocols for Mobile and Vehicular Ad Hoc Networks: A Comparative Analysis

Preetida Vinayakray-Jani[1*], Sugata Sanyal[2]

[1]DA_IICT, Gandhinagar, India,
preeti.vinayakray@gmail.com
[2]Tata Institute of Fundamental Research, Mumbai
sanyals@gmail.com

*Corresponding Author

*Abstract:* We present comparative analysis of MANET (Mobile Ad-Hoc Network) and VANET (Vehicular Ad-Hoc Network) routing protocols, in this paper. The analysis is based on various design factors. The traditional routing protocols of AODV (Ad hoc On-Demand Distance Vector), DSR (Dynamic Source Routing), and DSDV (Destination-Sequenced Distance-Vector) of MANET are utilizing node centric routing which leads to frequent breaking of routes, causing instability in routing. Usage of these protocols in high mobility environment like VANET may eventually cause many packets to drop. Route repairs and failures notification overheads increase significantly leading to low throughput and long delays. Such phenomenon is not suitable for Vehicular Ad hoc Networks (VANET) due to high mobility of nodes where network can be dense or sparse. Researchers have proposed various routing algorithms or mechanism for MANET and VANET. This paper describes the relevant protocols, associated algorithm and the strength and weakness of these routing protocols.

## I. INTRODUCTION

An emerging Mobile Ad hoc Networks (MANET) and Vehicular Mobile Networks (VANET) are expected to form network centric communications. Large number of mobile nodes communicates through single or multi-hop routing protocols. Although VANET is one of the classified scenarios of MANET, VANET nodes form highly dynamic network where node density could be either dense or sparse. Besides vehicle radios have very limited radio range and must communicate with one another by multi-hop routing protocols. Apparently, widely varying mobility characteristics of mobile or vehicular nodes are expected to have a significant impact on the performance of routing protocols. Therefore even though researchers have developed routing protocols like Ad hoc On-demand Vector (AODV), Dynamic Source Routing (DSR), Destination Sequence Distance Vector (DSDV) etc. for MANET [2], these protocols cannot be directly adopted in VANETs, efficiently, because of the rapid variation in link connectivity, high speed and extremely varied density of vehicular nodes in VANET. Researchers have developed special routing protocols for VANET [3], and these are aimed to adapt rapidly changing mobility pattern of the vehicular nodes.

Although such mobility characteristics exhibit spatial or temporal dependency of nodes, they are insufficient to capture some important mobility characteristics of scenarios in which MANETs may be deployed, i.e. the mobility characteristics generate protocol independent metrics [18]. But eventually this protocol independent metrics significantly influences the routing protocol performance. Attempt is made to categorize and summarize the routing protocols, as per the design factors, that influence the mobility performance.

This paper attempts to provide design factors that affect MANET and VANETs in section II. Subsections II also provide classification and qualitative comparison of MANET and VANET routing protocols. Finally section III discusses conclusion and open issues of developed or proposed routing protocols.

## II. DESIGN FACTORS THAT AFFECTS THE ROUTING PROTOCOLS

In general, routing protocols designed for MANET and VANET are categorized from topology point, these are either flat, hierarchical or position based; Communication paradigm (uni-cast or multicast or broadcast), Delay tolerance, Quality of service, Cluster based routing

### A. Topology

*Flat topology*: MANET routing protocols Optimized Link State Routing (OLSR), DSDV, Wireless Routing Protocol (WRP), Global State Routing (GSR), Fisheye State Routing (FSR), Source Tree Adaptive Routing (STAR) [7], Distance Routing Effect Algorithm for Mobility (DREAM) represents flat topology where route updates are periodically performed that constantly updates the network topology. This periodic updates are, regardless of network load, bandwidth or scalability. Such protocols are proactive and do not provide power saving as router updates are made periodically

Alternatively researchers have also developed there are reactive protocols in like AODV, Label-based Multipath Routing (LMR), Temporally-Ordered Routing Algorithm (TORA), Location Aided Routing (LAR), Zone Routing Protocol (ZRP), Flow Oriented Routing Protocol (FORP) where routing update is made on demand. In this type of protocol design active routes between sender and receiver nodes is determined by making *route discovery*. *Route discovery* is made by flooding network with *route request* and receiving *route response* packets in network. Such phenomena, helps nodes to conserve power as there are no periodic signals to respond.

These MANET protocols are not suitable for VANET, as discovering routing path is time consuming as vehicular node's speed is high.

*Hierarchical or Hybrid*: In MANET routing protocol like ZRP [4], represent this category that uses the hybrid approach to improve scalability of routing protocol. By considering proactive and reactive mechanisms, ZRP divides network in intra and inter zones, where intra-zone protocols are proactive and inter-zone protocols are reactive. Although this protocol improves scalability, lack of implementation feasibility makes this routing aspect unsuitable for VANET.

*Position-based:* Position based routing protocols uses location aided approach for MANET.

In VANET vehicular nodes either communicate with another vehicle (V2V) or road side vehicle (V2R). In existing infrastructure and ad hoc nodes of IEEE 802.11 wireless standard the time required to authenticate and associate with Basic Service Set (BSS) is too long to be considered by VANET. Therefore 802.11p standard will provide wireless devices with ability to communicate through short-duration messages, necessary to communicate between a high speed vehicle and a stationary roadside unit (V2R) including high speed vehicle (V2V). This mode of operation is known as Wireless Access in Vehicular Environment (WAVE) [19].

Although this V2V communication decentralized, it is robust and supports the low data transport times for emergency [10] warning, Such thing is not feasible with roadside cellular base station as they are often overwhelmed by calls in emergency, due to lack of load balancing mechanism to avoid congestion in network [5].

In MANET the proposed routing protocol LAR, uses information about location through geographic coordinates or relative position of nodes to generate route information thus by reducing overhead of traditional flooding mechanism. Moreover location service may be built into nodes or distributed location services may be utilized [12, 5].

The position based routing approach was designed for MANET routing protocol called Greedy Perimeter Stateless Routing (GPSR) [25]. In this greedy forwarding strategy is used to forward messages toward known destination. However if at one or multi hop, there are no nodes in direction of destination then it uses the perimeter mode. Usage of such routing strategy in VANET is not efficient as in urban area radio obstruction restricts the effective route and usage of perimeter mode is often required. During obstruction this perimeter mode uses the created planner graph that causes the message to be delivered immediate node instead of farthest reachable node. Thus more nodes will carry messages, eventually increasing delays. Such inefficiency can also cause messages to be delivered in wrong direction when node moves from communication range of one node to another.

As a result VANET uses Geographic Source Routing (GSR) [11]. This particular routing mechanism uses Dijkstra's shortest path algorithm to find shortest path between source and destination. Using static street map in piror and location information about each node, source forwards the message to destination and computes route to destination using Dijkstra's shortest path algorithm. The source message computes the sequence of intersection that must be traversed in order to reach destination. Although this algorithm is VANET specific it does not consider vehicle density, however authors acknowledges this and can see a potential to improve this routing mechanism.

Another position based routing protocol called Spatially Aware packet Routing (SAR) [16], tries to prevent limitations of recovery strategy used by GPSR of MANET. It is similar to GSR, but relies upon the external service such as Geographical Information Service (GIS) to extract street map and construct 'spatial model' to calculate shortest path to route packet to a destination. When shortest path is decided, unlike GSR, it determines the geographical locations that need to be travelled in embedded in packet header. When node needs to forward packet it uses this immobile physical location information to route packet to next geographic location. Thus it avoids the greedy strategy like GSR toward destination. Author does provide recovery strategy if forwarding node cannot find next location specified in packet header. In one method it suggests the usage of *suspension buffer* to store information till node finds suitable location. In another method node greedily forwards packet to destination. Usage of *suspension buffer* provides the high packet delivery ratio with expense of delay compare to no recovery strategy in SAR.

Unlike GSR, Anchor-based Street and Traffic (A-STAR) [20] aware routing uses bus routes to find routes with high probability for packet delivery. It uses the geographic forwarding points to route packet to destination, including route information to determine traffic density. However this static approach is less optimal compare to dynamic approach that utilizes latest traffic condition information.

In VANET, Connectivity Aware Routing (CAR) [15] maintains the cache of successful routes between various source and destination pairs. Nodes using CAR periodically sends HELLO beacons indicating the "velocity vector" information. On receiving this information receiving node will update its neighbour table and calculates its own velocity vector and velocity vectors of its neighbours. The entries in table expire after two HELLO intervals. However this HELLO beacon interval adapts as per traffic density, by increasing its frequency when traffic is sparse and by decreasing when traffic is dense. To maintain routing paths as the vehicle changes its position, *guards* are utilized to avoid the repetition of discovery phase of route. If the node at a route end point changes its direction then node activates *guard* with old and new velocity vectors. The node that is aware of a *guard* can use *guard* table information to ensure the delivery of messages to destination node that has moved. Once guard aware node receives a message addressed to the relocated node, it will add the guard coordinates as an anchor point to the message. Then it estimates the new position of the destination and forwards the message. Protocol also suggests two recovery strategies like "timeout algorithm with active cycle" and "walk around error recovery" to rectify the routing error incurred due to communication gaps between two anchor points or guards that are not maintained due to low traffic. Without making usage of map of location services, this protocol shows the ability to create the virtual infrastructure through '*guards*'. Protocols also provide the street and traffic awareness during discovery phase and maintains the route and adapts to traffic densities.

*B. Communication Paradigm*

In general communication paradigm include unicast, multicast, geocast, anycast, geographical anycast communication. Unicast communication provides one-to-one communication where target node location is known precisely or it is in the communication range through single or multi hop distance. Multicast or broadcast communication provides one-to-many communication where many single node can communicate with group of target nodes identified by common destination address. Multicasting is interpreted for group oriented communication. This type of communication paradigm is more suitable for applications that will require dissemination of messages to many different nodes in the network. The specialized form of multicast group is also called geocast where nodes are within particular geographic location relative to source able to receive geocast messages. In addition to this there is also another specialized form of multicast called anycast where a node sends message to any destination node in a group of nodes. This anycast also provides data acquisition feature where a nodes sends messages to certain geographic area to request data from any node found in that geographic location, called geographical anycast.

Many multicasting protocols have been proposed for MANET as well as for VANET. Designed and developed MANET based routing protocols are either using tree structure or mesh structure. Within the MANET working group at IETF two proposed multicast routing for ad hoc networks are Multicast Ad-hoc On-demand Distance Vector (MAODV) [26] and On-demand Multicast Routing Protocol (ODMRP) [27].

MAODV uses the shared bi-directional multicast tree while ODMRP maintains the tree topology. In MOADV with hard state of connected links, any link breakage force actions to repair the tree. Group leader in multicast tree maintains the up to date information about multicast tree by periodically sending group hello message and receiver unicasts the reply back to source. But if intermediate node on route path move away, the reply is lost, eventually route is lost.

Unlike MAODV, ODMRP being mesh topology, alternative path is feasible where link failure need not trigger the re-computation of the mesh. Any broken link eventually time out and route information for source and receiver is periodically refreshed by the source. The broadcasted route refreshes from every source could result in scalability issue if intermediate nodes are not part of multicast group, resulting in extra processing overhead. This makes tree based MAODV topology more efficient as it avoids sending duplicate packets to receivers. However in high mobility environment where topology changes very fast, tree-based MAODV is not suitable as unicast reply back to source is unable to reach if intermediate node in route path moves away. However in mesh based ODMRP alternative routes updates are broadcasted from receiver to source, making more robust against link failure with expense of associated overhead. Therefore compared to tree based topology, mesh based topology outperforms in high mobility environment.

## C. Delay Tolerance Network

Sparse MANETs are a class of ad hoc networks where node density is low and contacts between the nodes in network occurs infrequently. As a result, the network graph is rarely, if ever, connected where message delivery must be delay tolerant. However traditional MANET routing Protocols make the assumption that the network graph is fully connected and fail to route messages if there is no complete route from source to destination at the time of sending. For this reason traditional MANET routing protocol cannot be used in sparse MANETs. A key challenge is to find a route that can provide good delivery performance and low end-to-end delay in a disconnected graph where nodes may move freely

To overcome this issue, node mobility is exploited to physically carry messages between disconnected parts of network. The scheme that exploits the node mobility, referred to as mobility assisted routing that employs the *store-carry-and-forward* model is used. Mobility assisted routing consists of each node independently making forwarding decisions that take place when two nodes meet.

In VANET, when few vehicles are equipped with wireless transceivers, network will be sparse; delay tolerant routing algorithms are needed. The proposed Motion Vector Algorithm (MOVE) [8] for V2R VANET considers sparse network where prior prediction must be made for rare opportunistic routing. It is assumed that every node has knowledge of its own position and heading, where destination is a fixed globally known location. From this current vehicular node finds closest distance between vehicle and message destination along its trajectory. Current vehicular node periodically sends HELLO message. Neighbouring nodes sends RESPONSE message to make itself known to current vehicular node. Given the direction of where neighbouring node is heading; current node determines the shortest distance to destination along the trajectory of neighbouring node. The current node then makes decision to forward the message while determining the each vehicle's current distance from destination. This algorithm where data delivery rate is higher for sparse network, compared to greedy, position based routing and uses less system buffer space. With resulted performance evaluation, authors have noted that if routes are consistent and uniform, greedy position based routing performs better than MOVE.

In line with MOVE algorithm another algorithm called Scalable Knowledge based Vehicular Routing (SKVR) [1], also makes the usage of the predictable routes and vehicle schedules. It divides the network in inter-domain and intra-domain. In inter-domain routing source and destination belong to different routes whereas in intra-domain source and destination belong to same route. In inter-domain algorithm, message is forwarded to a vehicle travelling in destination domain and once destination domain is reached intra-domain message delivery procedure will be followed. In intra-domain messages are sent in forward or reverse directions, depending on the entires of contact list. If the sending vehicle contact list does not contain any vehicle in the destination's domain, then messages are delivered to the other vehicles in contact list. When vehicles along the same route encounter one another, a node carrying a message must decide whether to continue buffering the message, or to forward it, based on the direction information of the vehicle.

Using strategy called 'carry-and-forward' Vehicle Assisted Data Delivery (VADD) [17] algorithm allows packets to be carried by vehicles in sparse network and eventually relaying it to appropriate node when it enters in broadcasting range. Each node in VADD knows its own position and also requires external street map that includes traffic statistics. Selection of the candidate node, to which message need to be forwarded, is encountered through different selection criteria. However such criteria are either not scalable or consumes more bandwidth through duplication of packets. Authors have observed while using VADD, network becomes unstable as vehicle density decreased, because optimal paths were not available and because algorithm relies upon probabilistic traffic density information.

Unlike VADD, Static Node Assisted Adaptive Vehicular routing (SADV) [6] where static node has capability to store a message until it can forward the message to a node travelling on the optimal path. Algorithm also dynamically adapts to varying traffic densities in network, so that every node can measure the amount of time required to deliver message. However like any 'store-and-forward' this algorithm requires the efficient buffer management. By using 'Least Delay Increase' strategy, where static node checks which paths are currently available and eliminates packets which will not significantly increase their delivery delay.

Routing called Geographical Opportunistic (GeOpps) [9] routing in delay tolerant network is using opportunistic routing with carry-and-forward approach to route messages. Algorithm assumes that vehicle is using GPS and Navigation system that helps to route and locate static road site unit.

*D. Quality of Service (QoS)*

QoS routing strategy is not followed by any traditional MANET routing protocols. However there are research attempt to integrate such strategies within MANET routing protocols.

Multi-hop Routing Protocols for Urban VANET (MURU) [13], estimates quality factors of a route based on vehicle position, speed and trajectories. Based on this quality factors MURU introduces new metric called 'Expected Disconnection Degree' (EDD). Hence MURU nodes need to know its own position and have external street map including presence of efficient location service. This new metric value considered to be low as EDD, is an estimation of probability that determines the breakability of route during given time period. Based on destination location and street map, source node calculates the shortest trajectory to the destination to find route to destination. This shortest trajectory detail is stored in the packet and is used as a directional guideline for Route Request (RREQ) message. Node receiving RREQ message calculates EDD of the link between two subsequent nodes. MURU uses pruning method to improve the scalability of RREQ message, where node receiving RREQ message will wait for backoff delay that is directly proportional to the EDD between the previous forwarder of RREQ and current one. During this backoff interval the node determines whether to drop the RREQ message or rebroadcast it. Nevertheless, by using pruning method broadcasting area iteratively becomes smaller to receive RREQ broadcast. Eventually when destination receives the RREQ message from different routes, it selects the route with smallest EDD. This smaller broadcasting area is problematic if the next hop node is located outside of broadcasting range. However with low overhead and delay, MURU provide quality route with high percentage of throughput.

Another algorithm called Prediction Based Routing (PBR) [14], focussed on providing Internet connectivity to vehicles. This algorithm assumes that each vehicle has knowledge of its own position. The algorithm takes advantage of the less erratic vehicle movement patterns on road to predict the duration and expiry of a route from a client vehicle to a mobile gateway vehicle. Just before route failure is predicted, PBR pre-emptively seeks new route to avoid loss of service. However, it is unclear that how gateway will share bandwidth demand with number of vehicles.

*E. Clustering based routing*

Clustering is a process that divides the network into interconnected substructures, called structures. A group of nodes identifies themselves to be a part of cluster and a node designated as cluster head (CH) will broadcast the packet to cluster. The stability of node is the key to create the stable cluster infrastructure. There have been attempts to study cluster-based routing protocols in MANET. VANETs behave in a different way than the model that predominate in MANET's research, are due to driver behaviour, constraints on mobility and high speeds.

In MANET, Weighted Clustering Algorithm (WCA) [21] based on the use of weight metric that include several system parameters like the node-degree, distance with all its neighbours, node speed and time spent as a CH. Each node obtains the weight value of other nodes and CHs through re-broadcasting. As a result it induces overhead. If node moves into region which is not covered by CH, then once again cluster set-up process gets invoked. Such procedure is time consuming as it introduces more overhead to process. The performance of WCA is enhanced by algorithm called Distributed Weighted clustering Algorithm (DWCA) [22], which localizes the configuration and reconfiguration of cluster and restricts the power requirement on CHs.

In VANET, a reactive Location Routing Algorithm with Cluster Based Flooding (LORA-CBF) [23], where each node can be CH, gateway or cluster member. For each cluster there is CH, a node that connects two clusters called gateway. The packets are forwarded by protocol similar to greedy routing. If location of destination is not available then source will sent *location request*. This is similar to *route request* in AODV, but only CH and gateways can disseminates the *location request* and *location reply*. Performance results show the network mobility and size of the network affects the performance of AODV and DSR [2], more significantly than LORA-CBF.

Another VANET routing algorithm called Clustering for Open IVC Networks (COIN) [24], where CH is based on vehicular dynamics and driver intensions. Performance shows that COIN represents more stable clustering structure of VANET, at the cost of little overhead.

III. CONCLUSIONS AND OPEN ISSUES

In this paper attempt is made to provide comparative and qualitative analysis of MANET and VANET routing protocols by categorizing them within five different design factors.

Although foundation of MANET and VANET routing protocols is well established; it is essential to make comprehensive performance evaluation of various algorithms, by implementing them in real-time scenario.

The performance of routing protocols MANET and VANET depends significantly on the mobility models and the density of nodes. Therefore it is essential to design routing protocols specific to given mobility models.

REFERENCES

[1] S. Ahmed, S.S. Kanere, "SKVR: Scalable Knowledge-based Routing Architecture for Public Transport Networks", *Proceedings of the 3rd International Workshop on Vehicular Ad hoc Networks (VANET'06)*, ACM, New York, NY, USA, 2006, pp. 92-93.
[2] R. Bai, M. Singhal, "DOA: DSR over AODV routing for mobile ad hoc networks", *IEEE Transactions on Mobile Computing*, vol.5, no.10, Oct. 2006, pp.1403-1416.
[3] E. Fonseca, A. Festag, "A survey of existing approaches for secure ad hoc routing and their applicability to VANETS", NEC network laboratories, 28 pages, Version 1.1, March- 2006, pp. 1-28.
[4] Z. J. Haas, M. R. Pearlman, P. Samar, "The Zone Routing Protocol (ZRP) for Ad hoc Networks", IETF Internet Draft, July 2002. http://tools.ietf.org/id/draft-ietf-manet-zone-zrp-04.txt

[5] J. Bernsen, D. Manivannan, "Unicast Routing Protocols for vehicular Ad Hoc Networks: A critical comparison and classification," In Journal of Pervasive and Mobile Computing 5, 2009, Elsevier, pp. 1-18.

[6] Y. Ding, C. Wang, L. Xiao, "A static-node assisted adaptive routing protocol in vehicular networks" *Proceedings of the fourth ACM international workshop on Vehicular ad hoc networks, VANET'07*, ACM, New York, NY, USA, 2007, pp. 59-68.

[7] F. Giudici, E. Pagani, "Spatial and Traffic-Aware Routing (STAR) for Vehicular System." Proceedings of First International Conference on High Performance Computing and Communications, Lecture Notes in Computer Science. Publisher: Springer Berlin Heidelberg, Vol. 3726, 2005, pp. 77-86

[8] J. LeBrun, C.N. Chuah, D. Ghosal, M. Zhang, "Knowledge-based opportunistic forwarding in vehicular wireless ad hoc networks", *Proceedings of the 61st IEEE Vehicular Technology Conference (VTC)*, vol. 4, 30 May-1 June 2005, pp. 2289- 2293.

[9] I. Leontiadis, C. Mascolo, "GeOpps: Geographical Opportunistic Routing for Vehicular Networks", *IEEE International Symposium on World of Wireless, Mobile and Multimedia Networks*, 18-21 June 2007, Espoo, Finland, pp. 1-6.

[10] C. Lochert, H. Hartenstein, J. Tian, H. Fussler, D. Hermann, M. Mauve, "A routing strategy for vehicular ad hoc networks in city environments", *Proceedings of the IEEE Intelligent Vehicles Symposium*, 9-11 June, 2003, pp. 156-161.

[11] C. Lochert, M. Mauve, H. Fussler, H. Hartenstein, "Geographic routing in city scenarios", *ACM SIGMOBILE Mobile Computing and Communications Review*, vol. 9, No. 1, January, 2005, pp. 69-72.

[12] M. Mauve, J. Widmer, H. Hartenstein, "A survey on position-based routing in mobile ad hoc networks", *Journal of IEEE Network,* vol.15, no.6, Nov/Dec 2001, pp.30-39.

[13] Z. Mo, H. Zhu, K. Makki, N. Pissinou, "MURU: A Multi-hop routing protocol for urban vehicular ad hoc networks", *Proceedings of the Third IEEE Annual International Conference on Mobile and Ubiquitous Systems Workshops*, 17-21 July 2006, pp.1-8.

[14] V. Namboodiri, L. Gao, "Prediction-based routing for vehicular ad hoc networks", *IEEE Transactions on Vehicular Technology*, Vol. 56, No. 4, July 2007, pp. 2332-2345 .

[15] V. Naumov, T. R. Gross, "Connectivity-aware routing (CAR) in vehicular ad-hoc networks", *Proceedings of the 26th IEEE International Conference on Computer Communications*, 6-12 May 2007, pp.1919-1927.

[16] J. Tian, L. Han, K. Rothermel, "Spatially aware packet routing for mobile ad hoc inter-vehicle radio networks", *Proceeding of the IEEE Intelligent Transportation Systems*, Vol. 2, 12-15 October, 2003, pp. 1546- 1551.

[17] J. Zhao, G. Cao, "VADD: Vehicle-Assisted Data Delivery in vehicular ad hoc networks", *Proceedings of the 25th IEEE International Conference on Computer Communications (INFOCOM),* 2006, pp. 1-12.

[18] Bhavyesh Divecha, Ajith Abraham, Crina Grosan and Sugata Sanyal "Analysis of Dynamic Source Routing and Destination-Sequenced Distance-Vector Protocols for Different Mobility models", *First Asia International Conference on Modelling and Simulation*, 27-30 March, 2007, Phuket, Thailand, pp. 224-229.

[19] Status of Project IEEE 802.11 task group p, "Wireless Access in Vehicular Environment" (WAVE)",
http://www.ieee802.org/11/Reports/tgp_update.htm

[20] R.C. Seet, G. Liu, B.S. Lee, C.H Foh, K.J. Wong and K.K. Lee, "A-STAR: A mobile ad hoc routing strategy for metropolis vehicular Communication", *Performance of Computer and Communication Networks*, Lecture notes in Computer Science, Vol. 3042, Publisher: Springer Berlin Heidelberg, 2004, pp. 989-999.

[21] M Chatterjee, S. Das and D. Turgut, "WCA: A Weighted Clustering Algorithm for Mobile Ad hoc Networks", Journal of Cluster Computing (Special Issue on Mobile Ad Hoc Networks), Vol. 5, No. 2, Springer, pp. 193-202, 2002

[22] W. Choi and M. Woo, "A Distributed Weighted Clustering Algorithm for Mobile Ad Hoc Networks", In Proceedings of IEEE Advanced International Conference on Telecommunications and International Conference on Internet and Web Applications and Services, 2006, pp. 73.

[23] R. A. Santos, A. Edwards, R.M. Edwards and N.L.Seed, "Performance evaluation of routing Protocols in Vehicular Ad Hoc Networks," The International Journal of Ad Hoc and Ubiquitous Computing, Vol. 1, No. 1, pp. 80-91, Inderscience, 2005

[24] J. Blum, A. Eskandarian and L. Hoffman, "Mobility Management in IVC Networks," In Proceedings of IEEE Intelligent Vehicles Symposium, 2003, pp. 150-155.

[25] B. Karp, H.T. Kung, "GPSR: Greedy Perimeter Stateless Routing for Wireless Networks," In Proceedings of the 6[th] International Conference on Mobile Computing and Networking, MobiCom'00, ACM, New York, NY, USA, 2000, pp. 243-254.

[26] E. M Royer, C. E. Perkins, "Multicast Ad hoc On-demand Distance Vector (MAODV) Routing," Internet Draft,
http://tools.ietf.org/html/draft-ietf-manet-maodv-00

[27] Y. Yi, Sung-Ju Lee, W. Su, M. Gerla, "On-demand Multicast Routing Protocol (ODMRP) for Ad hoc Networks", Internet Draft,
http://tools.ietf.org/html/draft-ietf-manet-odmrp-04